Data Mining

# Classifying ancient texts by inter-word spacing

In the 1940s, Claude Shannon applied statistical concepts to both written text and genetics, pioneering the field of information theory[1]. The seminal paper that exploited Shannon's concept of information entropy for the study of DNA sequences was by Mantegna *et al*[2]. Due to the mapping of the human genome, there is renewed interest in statistical techniques for classification and data mining of DNA sequences[3]. Given that DNA sequences can be viewed as possessing a 4 letter alphabet {A, C, G, T} it is not surprising that data mining and classification of text has inspired techniques for DNA and *vice versa*. Recent advances in these areas have been demonstrated by Ortuño *et al*[4] and Benedetto *et al*[5], for example.

Ortuño *et al*[4] have explored both written texts and DNA using a powerful new idea of inter-word spacing for extracting keywords. The inter-word spacing is defined as the word count between a word and the next occurrence of the same word in a text. All the inter-word spaces, for each case, are then counted up and the standard deviation is computed. This is then repeated for different words – the words are then ranked according to the standard deviation values, the highest first. The standard deviation $\sigma$ is then plotted versus the logarithm of the rank. Ortuño *et al*[4] have found that words with the highest $\sigma$ ranking tend to make better search engine keywords, as opposed to words with high hit counts.

We demonstrate a striking new result; in Figure 1, where standard deviation versus log of rank is plotted for *Koine* Greek source texts of the New Testament[6]. For simplicity only the books of *Luke*, *Matthew* and *Acts* are plotted here. The close match between the curves for *Acts* and *Luke*, compared to other books, appears to add weight to what has always been accepted by scholars; namely that the author of *Luke* was identical to that of *Acts*.

To check the significance of this match we introduce the idea of a quantitative measure obtained by comparing variances in spacing between words common to all the texts under examination. We use a two-distribution $\chi^2$ measure on the variances $\sigma^2$ of top-ranked keywords; the $\chi^2$ values for pairs of texts are put into the table below, with a lower $\chi^2$ indicating a closer match. This gives the following result for the gospels and *Acts*.

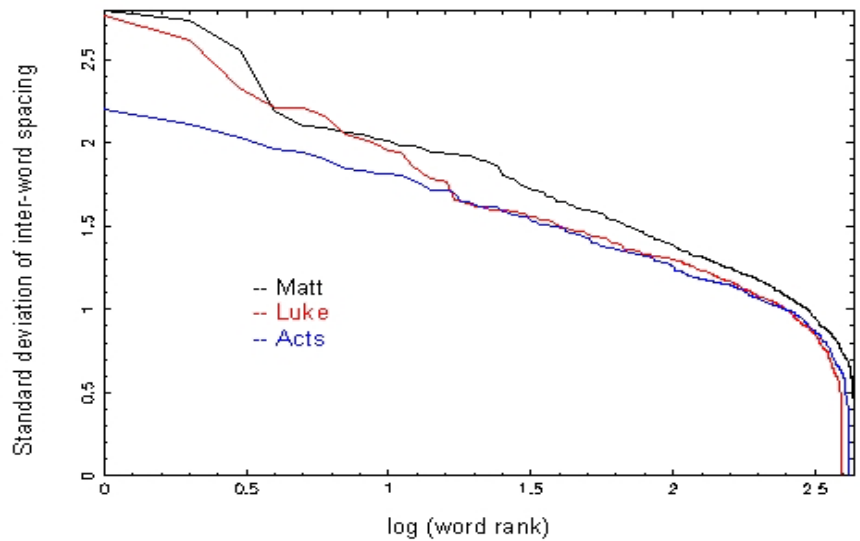

**Figure 1** The standard deviation $\sigma$ of the inter-word spacing (y-axis) for each word is ranked in descending order on a logarithmic scale (x-axis). Using the original *Koine* Greek text, a remarkably close match is obtained between the book of *Luke* and the book of *Acts* in the New Testament, which were written by the same author. For reference, a curve of a different author is shown (book of *Matthew*) illustrating a distinct separation (this is the upper curve). Although the match between *Luke* & *Acts* deviates for a log rank <1.2, this represents less than 4% of the total curve (due to the logarithmic scale). Note that uncommon words occurring less than 5 times in each text are not included in the ranking, as their $\sigma$ values are not significant.

|       | Matt | Mark | Luke | John | Acts |
|-------|------|------|------|------|------|
| Matt  | 0.00 | 3.91 | 2.20 | 6.05 | 3.95 |
| Mark  | 3.91 | 0.00 | 3.21 | 5.53 | 4.90 |
| Luke  | 2.20 | 3.21 | 0.00 | 2.42 | 2.02 |
| John  | 6.05 | 5.53 | 2.42 | 0.00 | 3.17 |
| Acts  | 3.95 | 4.90 | 2.02 | 3.17 | 0.00 |

As a check against a known benchmark, the following table compares works by Charles Dickens (*Great Expectations* and *Barnaby Rudge*) and Thomas Hardy (*Jude the Obscure* and *Tess of the d'Urbervilles*):

|      | Jude | Tess | Barn | GE   |
|------|------|------|------|------|
| Jude | 0.00 | 1.05 | 4.92 | 8.24 |
| Tess | 1.05 | 0.00 | 2.34 | 4.59 |
| Barn | 4.92 | 2.34 | 0.00 | 1.86 |
| GE   | 8.24 | 4.59 | 1.86 | 0.00 |

As expected, lowest $\chi^2$ scores are obtained for the correct author match.

A source of considerable historical debate has been the question of authorship of the book of *Hebrews* in the New Testament. An open question would be to apply our technique to possibly eliminate some of the conjectured authors.

In conclusion, our results add weight to the generally accepted hypothesis of a common author between the books of *Luke* and *Acts*. Future developments in this area may shed some light on a number of historical debates surrounding the question of authorship. Applying these types of tests to DNA may be of interest in the study of phylogenic relationships.

**Matthew J. Berryman,**
**Andrew Allison, Derek Abbott**
*Centre for Biomedical Engineering,
Department of Electronic and Electrical
Engineering, University of Adelaide,
Adelaide, SA 5005, Australia
e-mail: dabbott@eleceng.adelaide.edu.au*